\documentclass[pra,twocolumn]{revtex4}
\begin{document}
\title{Character of Locally Inequivalent Classes of States and Entropy of Entanglement}
\author{Indrani Chattopadhyay\protect\(^{1}\protect \)}

\affiliation{School Of Science, Netaji Subhas Open University, K-2,
Fire Station, Sector-V, Salt Lake, Kolkata-700 091, India}

\author{Debasis Sarkar\protect\(^{2}\protect \)}

\affiliation{Department of Applied Mathematics, University of
Calcutta, 92, A.P.C. Road, Kolkata- 700009, India}

\begin{abstract}
In this letter we have established the physical character of pure bipartite states
with the same amount of entanglement in the same Schmidt rank that either they are
local unitarily connected or they are incomparable. There exist infinite number of
deterministically locally inequivalent classes of pure bipartite states in the same
Schmidt rank (starting from three) having same amount of entanglement. Further, if
there exists incomparable states with same entanglement in higher Schmidt ranks
(greater than three), then they should differ in at least three Schmidt coefficients.

PACS number(s): 03.67.Hk, 03.65.Ud.


\end{abstract}

\maketitle

Quantum mechanics shows many counterintuitive properties in physical
systems. However, these peculiar characteristics turn out as the
fundamental features of different quantum systems. Entanglement is
one of the most striking phenomena in recent times. Apart from its
peculiar characteristics, entanglement is considered as an important
physical resource in various quantum information processing tasks
\cite{mes,densecoding,upbBennett}. Thus the characterization and
quantification of entanglement would always be one of the major
issue to understand the behavior of composite quantum systems. It is
almost convincing that the quantification of pure state entanglement
in bipartite system through the von-Neumann entropy of reduced
density matrices is complete. Popescu and Rohrlich showed that
\cite{UniqueMeasure} von-Neumann entropy quantifies the unique
measure of pure state entanglement in bipartite level. For pure
bipartite states, both the entanglement of formation and distillable
entanglement \cite{ErrorCorr} are the same with the entropy of
entanglement and every measures of entanglement that should satisfy
some fundamental criteria, would necessarily collapse with the
entropy of entanglement. Thus, it is assumed that the entropy of
entanglement should necessarily reflect all the possible non-local
features of pure entangled states in bipartite level. However, in
this paper our findings are something beyond the entropy of
entanglement. It generates with the connection between entanglement
and LOCC. For pure bipartite states all states which are connected
by local unitary operations, have the same amount of entanglement
and have also the same set of Schmidt coefficients. But the general
character of pure bipartite entangled states with the same amount of
entanglement in the same Schmidt rank is not completely known to us.
Also, if there exist such states with different Schmidt
coefficients, then what is the physical nature of such states? In
this letter, we would able to answer the above problems through the
existence of incomparable states in pure bipartite level.

The notion of incomparable states in pure bipartite level is due to
Nielsen \cite{nielsen} through the majorization criteria for
deterministic conversion of pure entangled states under LOCC.
Considerable amount of effort has been spent with the possibility
and impossibility of manipulating pure entanglement under
deterministic or stochastic LOCC
\cite{LoPopescuVidal,catalysis-multicopy,Ishizaka}. However, the
character of locally inequivalent, i.e., incomparable states is not
clearly understood at least in pure bipartite level. It is peculiar
that although there is no restriction from the amount of
entanglement contained in a pure bipartite state, but it is
impossible to convert this state to another lower entangled state,
if they are incomparable to each other. To understand the basic
nature of such states, recently it is found that there are some
deeper relations between the existence of incomparable states and
different no-go theorems and incomparability may be used as a
detector of unphysical operations \cite{impos}. The important factor
we would like to explore in this work is the existence of infinite
number of pure bipartite entangled states with the same entanglement
but all are incomparable to each other, i.e., there exists infinite
number of pure entangled states having different Schmidt
coefficients with the same entanglement in the same Schmidt rank. It
starts from $3\times 3$ systems. The nature of such states in higher
dimensional systems are also quite surprising. They should differ in
at least three Schmidt coefficients and up to that level equivalent
to the case of incomparable states in $3\times 3$ systems with the
same amount of entanglement. Our proof is analytical, supported by
numerical results and it would necessarily reflect the basic nature
of entanglement as a non-local feature beyond the entropy of
entanglement.

Let us first investigate the possible relations between entanglement
of two pure bipartite states in $3\times 3$ systems. Suppose,
$\alpha_1, \alpha_2, \alpha_3$ and $\beta_1, \beta_2, \beta_3 $ are
the Schmidt coefficients of two pure bipartite state $|\Psi\rangle$
and $|\Phi\rangle$ of Schmidt rank three, i.e., $1> \alpha_i ,
\beta_i > 0  ~~\forall ~i=1,2,3$ and $\alpha_1 + \alpha_2 + \alpha_3 
= 1= \beta_1  + \beta_2 + \beta_3$. Then entanglement of the pure states 
are given by the von-Neumann entropy of the reduced density matrices as,
\begin{equation}
\begin{array}{lcl}
E(|\Psi\rangle)= -(\alpha_1\log_2{\alpha_1} + \alpha_2 \log_2{\alpha_2}+ \alpha_3 \log_2{\alpha_3}),\\
E(|\Phi\rangle)= -(\beta_1\log_2{\beta_1} + \beta_2 \log_2{\beta_2}+
\beta_3 \log_2{\beta_3})
\end{array}
\end{equation}
\textbf{Theorem-1:} \emph{Let $|\Psi\rangle, |\Phi\rangle$ be two
pure bipartite states with Schmidt rank three, having same amount of
entanglement. If one pair of Schmidt coefficient for the two pure
states are equal ($\alpha_i = \beta_i$ for some $i=1,2,3$), then so
also for the other two Schmidt coefficients, i.e., $\alpha_i =
\beta_i,~ \forall~i=1,2,3$.}

\emph{Proof.} Let $\alpha_1 = \beta_1$, i.e., the largest Schmidt coefficients of
the pure states $|\Psi\rangle, |\Phi\rangle$ be equal. Also, we
assume that $E(|\Psi\rangle)= E(|\Phi\rangle)$. Then, $\alpha_2 +
\alpha_3 =1-\alpha_1=1-\beta_1=\beta_2 + \beta_3$, Thus we may
construct two pure bipartite states of Schmidt rank two, as
\begin{equation}
\begin{array}{lcl}
|\Psi'\rangle\equiv (\alpha, 1-\alpha), ~ |\Phi'\rangle\equiv (\beta, 1-\beta)
\end{array}
\end{equation}
where $\alpha=\frac{\alpha_2}{\alpha_1}, ~
\beta=\frac{\beta_2}{\beta_1}$. Then using $E(|\Psi\rangle)=
E(|\Phi\rangle)$ we observe that $E(|\Psi'\rangle)= E(|\Phi'\rangle),$
which would necessarily imply that $|\Psi'\rangle$ and $|\Phi'\rangle$
have the same set of Schmidt coefficients, i.e., $\alpha = \beta$. Thus,
$\alpha_2 = \beta_2$ and $\alpha_3 = \beta_3$. So, the pure states
$|\Psi\rangle, |\Phi\rangle$ must necessarily have the same set of Schmidt
coefficients.

In a similar manner, if either of $\alpha_2 =
\beta_2$ or $\alpha_3 = \beta_3$, then it is also be the case that the pure
bipartite states $|\Psi\rangle, |\Phi\rangle$ must have the same
set of Schmidt coefficients.

Before going to describe the next result, we recall the notion of
incomparable states in pure bipartite level \cite{nielsen}. Two pure
bipartite states $|\Psi\rangle$ and $|\Phi\rangle$ of $m\times
n$ system with $\min\{m,n\}\leq d$ are said to be comparable to each
other if and only if the Schmidt coefficients
$\alpha_1,\alpha_2,\cdots,\alpha_d,$ and
$\beta_1,\beta_2,\cdots,\beta_d$ corresponding to the states
$|\Psi\rangle$ and $|\Phi\rangle$ should satisfy the following
relations,
\begin{equation}
\begin{array}{lcl}\sum_{i=1}^{k}\alpha_{i}\leq
\sum_{i=1}^{k}\beta_{i},~ ~\forall~ ~k=1,2,\cdots,d
\end{array}
\end{equation}
where $\alpha_{i}\geq \alpha_{i+1}\geq 0$ and $\beta_{i}\geq
\beta_{i+1}\geq0,$ for $i=1,2,\cdots,d-1,$ and $\sum_{i=1}^{d}
\alpha_{i} = 1 = \sum_{i=1}^{d} \beta_{i}$.
It is known as majorization \cite{maj} criteria of two vectors
formed by the Schmidt coefficients of the states and it provides us
the necessary and sufficient condition for converting $|\Psi\rangle$
to $|\Phi\rangle$ under deterministic LOCC. As a consequence of
non-increase of entanglement by LOCC, if $|\Psi\rangle\rightarrow
|\Phi\rangle$ is possible under LOCC with certainty, then
$E(|\Psi\rangle)\geq E(|\Phi\rangle)$ where,
\begin{equation}
\begin{array}{lcl}
E(|\Psi\rangle)= -\sum_{i=1}^d \alpha_i\log_2{\alpha_i} ,\\
E(|\Phi\rangle)= -\sum_{i=1}^d \beta_i\log_2{\beta_i}.
\end{array}
\end{equation}
If the above criterion [eqn. (3)] does
not hold, then we usually denote it by $|\Psi\rangle\not\rightarrow
|\Phi\rangle$ and if both $|\Psi\rangle\not\rightarrow |\Phi\rangle$
and $|\Phi\rangle\not\rightarrow |\Psi\rangle$ occur, then we denote
it as $|\Psi\rangle\not\leftrightarrow |\Phi\rangle$ and call
$(|\Psi\rangle, |\Phi\rangle)$ as a pair of incomparable states
\cite{nielsen,catalysis-multicopy}. For $3\times 3$ states $|\Psi\rangle, |\Phi\rangle$
with Schmidt coefficients $\alpha_1, \alpha_2, \alpha_3$ and
$\beta_1, \beta_2, \beta_3 $ in decreasing order, the condition for
incomparability can be written in the simplified form
\begin{equation}
\begin{array}{lcl}\verb"either," ~ ~ ~ ~\alpha_1 > \beta_1 ~ ~\verb"and" ~ ~\alpha_3 > \beta_3\\

\verb"or," ~ ~ ~ ~ ~ ~ ~ ~ ~ ~\alpha_1 < \beta_1 ~ ~\verb"and" ~ ~\alpha_3 < \beta_3.
\end{array}
\end{equation}
So, if there exists two incomparable states in $3 \times 3$ system with the same amount of
entanglement, then all the Schmidt coefficients must be different.

\textbf{Theorem-2:} \emph{The amount of entanglement of any two
comparable, pure, bipartite states of $d\times d, ~ d \geq 3$ systems with different Schmidt
coefficients must necessarily be different.}

In other words, for any two Schmidt rank $d ~(\geq 3)$ states
$|\Psi\rangle$ and $|\Phi\rangle$ with different Schmidt coefficients,
\begin{equation}
\begin{array}{lcl}
|\Psi\rangle \rightarrow |\Phi\rangle \Rightarrow E(|\Psi\rangle)
> E(|\Phi\rangle)
\end{array}
\end{equation}

To prove the theorem, we use the concept of Schur Convexity and
its connection with the majorization of vectors.

\emph{Schur Convex Function:}\cite{Nielsen-thesis,maj,Schur-convex} A function 
$F: I^n \rightarrow \Re$ is called Schur Convex if,
\begin{equation}
\begin{array}{lcl}
x \prec y \Longrightarrow F(x) \leq  F(y), ~~~\forall ~x, y \in I^n
\end{array}
\end{equation}
where $I \subset \Re$, $\Re$ is the set of all real numbers
and $x \prec y$ means $x$ is majorized by $y$.

The function $F(x)\equiv F(x_1, x_2, \cdots, x_n), x_i \in I, \forall i= 1,2, \ldots, n,$ 
is called Strictly Schur Convex, if and only if, the above inequality is strict
for all $x \in I^n$. All Schur convex functions are symmetric in
nature, i.e., invariant under any permutation, but the converse is
not true \cite{Nielsen-thesis,maj,Schur-convex}.

Also, a function $F: I^n \rightarrow \Re$ where $I \subset \Re$ is called Schur Concave(strictly) 
if and only if the function $F'=-F$ is  Schur Convex(strictly).

\textbf{Lemma}\cite{Nielsen-thesis,maj,Schur-convex}: Suppose a function 
$F: I^n \rightarrow \Re$ where $I \subset \Re$, is symmetric and have continuous partial
derivatives on $I^n$. Then $F(\cdot)$ is Schur Convex, if and only if,
\begin{equation}
\begin{array}{lcl}
(x_i-x_j)(\frac{\partial F}{\partial x_i}-\frac{\partial F}{\partial
x_j})\geq 0,~\forall ~x_i,x_j \in I ; i,j=1,2,\cdots,n
\end{array}
\end{equation}
It is strictly Schur convex if and only if the above
inequality is strict for all $x_i\neq x_j$.

The well known example of strict Schur concave function is the Shannon entropy
of a probability distribution, i.e., $H(p) = -\sum_{i=1}^{n} p_i \log_2 p_i, ~
0\leq p_i \leq 1, \sum_{i=1}^{n} p_i =1$ \cite{Nielsen-thesis}.

Now, consider a pure bipartite state $|\Psi \rangle$ of
Schmidt rank $d$ with Schmidt vector
$\lambda_{|\Psi\rangle}=(\lambda_1,\lambda_2,\cdots,\lambda_d)$. Then,
$E(|\Psi\rangle)=-\sum_{i=1}^{d} \lambda_i \log_2 \lambda_i,$ where
$0\leq \lambda_i \leq 1,$ $\sum_{i=1}^{d} \lambda_i =1$. It is easy to check that,
\begin{equation}
\begin{array}{lcl}
(\lambda_i-\lambda_j)(\frac{\partial E}{\partial
\lambda_i}-\frac{\partial E}{\partial
\lambda_j})\\
=(\lambda_i-\lambda_j)\{(-\log_2
\lambda_i- \log_2 e)-(-\log_2
\lambda_j-\log_2 e)\}\\
=(\lambda_i-\lambda_j)\{\log_2(\frac{\lambda_j}{\lambda_i})\}
<0,~~~\forall~\lambda_i\neq \lambda_j
\end{array}
\end{equation}
So from the lemma, we conclude that $E: I^d \rightarrow \Re$ where $I
= [0,1]$ is a strictly Schur Concave function.

Proof of Theorem-2: Suppose, $|\Psi\rangle$, $|\Phi\rangle$ are any two pure
bipartite states of Schmidt rank $d, ~d\geq 3$. Then
$|\Psi\rangle\longrightarrow|\Phi\rangle$ under deterministic LOCC
if and only if $\lambda_{|\Psi\rangle}\prec\lambda_{|\Phi\rangle}$,
where $\lambda_{|\Psi\rangle},\lambda_{|\Phi\rangle}$ are Schmidt vectors
of $|\Psi\rangle$ and $|\Phi\rangle$ \cite{nielsen}. Now, for different Schmidt
vectors of $|\Psi\rangle$ and $|\Phi\rangle$, the strict Schur concavity of
the function $E$ implies,
\begin{equation}
\begin{array}{lcl}
~~~~~~~ -E(\lambda_{|\Psi\rangle}) &<& -E(\lambda_{|\Phi\rangle}) \\
\Rightarrow ~~~~~~~E(\lambda_{|\Psi\rangle}) &>& E(\lambda_{|\Phi\rangle}). \\
\end{array}
\end{equation}

One may also check the results by algebraic method by considering the
different possible cases of majorization relation. We have mentioned
a case explicitly in the appendix. The above result is also true for
$d=2.$

Thus, we may conclude that the amount of entanglement of any two
comparable, $d\times d$ pure bipartite states with different Schmidt
coefficients, must necessarily be different.
This result is quite compatible with our natural intuition. However,
it would not imply immediately that all the pure bipartite states
with same entanglement are locally unitarily connected, or at least
locally connected. In contrary, there are infinite number of states
with the same entanglement even in the lowest possible dimension,
i.e., in $3\times 3$ systems, but incomparable in nature. The
theorems 1 and 2 readily imply that if there exist pure entangled
states with the same entanglement but different Schmidt coefficients
in $d\times d, ~~(d\geq 3)$ system, then they must be incomparable to each other.
The following example is a way how one could find such states numerically.

\textbf{Example:} Firstly, consider the pure bipartite state $|\Psi\rangle$,
with Schmidt coefficients  .45,.39,.16. The entropy of entanglement of the
state is $E(|\Psi\rangle)\approx 1.471215431$. Next, consider the following 
pair of states represented by their Schmidt vectors,
\begin{equation}
\begin{array}{lcl}
|\Phi_1\rangle &\equiv & (.49, .33676028, .17323972),\\
|\Phi_2\rangle &\equiv & (.49, .33676030, .17323970)
\end{array}
\end{equation}
Both the states $|\Phi_1\rangle$ and $|\Phi_2\rangle$ are
incomparable with $|\Psi\rangle$ and have the entanglement
$E(|\Phi_1\rangle)\approx 1.471215442, ~ E(|\Phi_2\rangle)\approx
1.471215423$. Then, upto the $8$ significant digits,
$E(|\Psi\rangle)=E(|\Phi_1\rangle)=E(|\Phi_2\rangle)= 1.4712154$ and
precisely, $E(|\Phi_1\rangle) > E(|\Psi\rangle) >
E(|\Phi_2\rangle)$. From the continuity of the von-Neumann entropy
function on Schmidt coefficients of the states, we may conclude that
there exist a pure $3\times 3$ state $|\Phi\rangle \equiv (.49,
.33676028 + \delta, .17323972 - \delta)$ where $0 < \delta <
.00000002$ between the states $|\Phi_1\rangle$ and $|\Phi_2\rangle$,
such that $E(|\Psi\rangle)=E(|\Phi\rangle)$ exactly. By construction  the
states $|\Psi\rangle$ and $|\Phi\rangle$ are incomparable with each
other. Now, further, if we look at the largest Schmidt coefficients .45 and
.49 of the states $|\Psi\rangle$ and $|\Phi\rangle$ respectively, we
observe that they are widely separated. The states have certain
distance with respect to Schmidt coefficients and if we consider
the largest Schmidt coefficient any value between .45 and .49, then we
could always find a state $|\Phi^{\prime}\rangle$ that has the same
amount of entanglement with $|\Psi\rangle$ and incomparable with it.
Thus, there are infinite number of pure bipartite states which have
same amount of entanglement, but incomparable to each other.

The character of any pair of pure bipartite states in $d\times d$
system with equal entanglement have a nice relation with the lower
dimensional incomparable states and ultimately we find that any pair
incomparable states should differ in at least three Schmidt coefficients.
Consider a pair of pure bipartite states in $d\times d$ system
having the Schmidt vectors,
\begin{equation}
\begin{array}{lcl}
|\Psi\rangle &\equiv& (\alpha_1, \alpha_2, \cdots \alpha_d),\\
|\Phi\rangle &\equiv& (\beta_1, \beta_2, \cdots \beta_d)
\end{array}
\end{equation} If they have the same entanglement, then either $\alpha_i ~= ~ \beta_i,~~\forall~~i=
1,2,\cdots,d,$ or, $\alpha_i ~\neq ~ \beta_i$ for at least $3$
values of $i\in \{1,2,\cdots,d\}$, i.e., they are either locally
unitarily connected or they are incomparable with at least three
different Schmidt coefficients. To be precise, if there is exactly
$k\leq (d-3)$ number of values of $i\in \{1,2,\cdots,d\}$ for which
$\alpha_i ~=~ \beta_i$, then there exists a pair of incomparable
pure bipartite states with Schmidt rank $d-k$, having same amount of
entanglement for which all the other $d-k$ Schmidt coefficients are
different. We would now show it for $k=1$. Suppose, the $j^{th}$
Schmidt coefficients of $|\Psi\rangle$ and $|\Phi\rangle$ are equal,
i.e., $\alpha_j ~= ~ \beta_j~=~ \kappa$(say) and $\alpha_i ~\neq ~
\beta_i$ for all other $i$. Then we may construct two pure bipartite
states with Schmidt rank $d-1$ as follows,
\begin{equation}
\begin{array}{lcl}
|\Upsilon\rangle &=& (\chi_1, \chi_2, \cdots \chi_{d-1}),\\
|\Omega\rangle &=& (\eta_1, \eta_2, \cdots \eta_{d-1})
\end{array}
\end{equation} with $\chi_i=\frac{\alpha_i}{1-\kappa}$ and $\eta_i=\frac{\beta_i}{1-\kappa}$
for $1\leq i<j$ and $\chi_i=\frac{\alpha_{i+1}}{1-\kappa}$ and
$\eta_i=\frac{\beta_{i+1}}{1-\kappa}$ for $j \leq i \leq {d-1}$.
Now,

$E(|\Psi\rangle) =  E(|\Phi\rangle$

$ \Rightarrow -\sum_{i=1}^d \alpha_i\log_2{\alpha_i}= -\sum_{i=1}^d
\beta_i\log_2{\beta_i}$

$\Rightarrow \sum_{i=1}^{j-1} \alpha_i\log_2{\alpha_i}+
\sum_{i=j+1}^{d} \alpha_i\log_2{\alpha_i}$

$~~~=  \sum_{i=1}^{j-1} \beta_i\log_2{\beta_i}+
\sum_{i=j+1}^{d}\beta_i\log_2{\beta_i}$

$\Rightarrow \sum_{i=1}^{j-1}
\frac{\alpha_i}{1-\kappa}\log_2{\frac{\alpha_i}{1-\kappa}}+
\sum_{i=j+1}^{d}
\frac{\alpha_i}{1-\kappa}\log_2{\frac{\alpha_i}{1-\kappa}}$

$~~~= \sum_{i=1}^{j-1}
\frac{\beta_i}{1-\kappa}\log_2{\frac{\beta_i}{1-\kappa}}+
\sum_{i=j+1}^{d}\frac{\beta_i}{1-\kappa}\log_2{\frac{\beta_i}{1-\kappa}}$

$\Rightarrow \sum_{i=1}^{j-1}
\frac{\alpha_i}{1-\kappa}\log_2{\frac{\alpha_i}{1-\kappa}}+
\sum_{i=j}^{d-1}
\frac{\alpha_{i+1}}{1-\kappa}\log_2{\frac{\alpha_{i+1}}{1-\kappa}}$

$~~~= \sum_{i=1}^{j-1}
\frac{\beta_i}{1-\kappa}\log_2{\frac{\beta_i}{1-\kappa}}+
\sum_{i=j}^{d-1}\frac{\beta_{i+1}}{1-\kappa}\log_2{\frac{\beta_{i+1}}{1-\kappa}}$

$\Rightarrow -\{\sum_{i=1}^{d-1}\chi_i\log_2{\chi_i}\}
= -\{\sum_{i=1}^{d-1} \eta_i\log_2{\eta_i}\}$

$\Rightarrow E(|\Upsilon\rangle) =  E(|\Omega\rangle).$

Also $(|\Psi\rangle, |\Phi\rangle)$ are incomparable imply, either
$\alpha_1\leq \beta_1$ and $\sum_{i=1}^m \alpha_i > \sum_{i=1}^m
\beta_i$ for some $m \in \{2,3,\cdots,d-1\}$ or $\beta_1\leq
\alpha_1$ and $\sum_{i=1}^n \beta_i >\sum_{i=1}^n \alpha_i$ for some
$n \in \{2,3,\cdots,d-1\}$. Then, either $\chi_1\leq \eta_1$ and
$\sum_{i=1}^m \chi_i > \sum_{i=1}^m \eta_i$ for some $m \in
\{2,3,\cdots,d-2\}$ or $\eta_1\leq \chi_1$ and $\sum_{i=1}^n
\eta_i>\sum_{i=1}^n \chi_i$ for some $n \in \{2,3,\cdots,d-2\}$,
i.e., $(|\Upsilon\rangle,|\Omega\rangle)$ are incomparable.

Proceeding in the same way, it is always possible to construct a
lower dimensional incomparable pair of states with same entanglement
from an upper dimensional one and they should differ in at least three Schmidt coefficients.

The above results have some immediate consequences in quantum information theory.
If someone is restricted to use non-maximally pure bipartite entangled states as
teleportation channel, then the optimal fidelity \cite{opttele} for sending
qudits are different for a pair of incomparable states with equal entanglement.
Thus, the capacity as channel is not always equal, however, same resource in
the sense of amount of entanglement, are used for the task. Again, if we mix
such incomparable states with simply identity (garbage state), then sometimes
they are PPT states and sometimes they are NPT states with the same mixing
probabilities but differ only with the incomparable states considered. Also,
the behavior of these locally inequivalent classes of states are different, if
we consider different measures of correlations \cite{preparation}. In particular,
it easy to check the non-monotonicity of concurrence with entanglement of formation \cite{commu}.

In conclusion we have found the physical character of pure bipartite states with the
same amount of entanglement in the same Schmidt rank. The number of such incomparable
states are infinite and the higher Schmidt rank incomparable states with equal
entanglement must differ with at least three Schmidt coefficients. Thus it is observed
that the entropy of entanglement is not always able to characterize the non-local
properties of pure bipartite states. The relations between locally inequivalent
pure bipartite states and different entanglement measures will be the important
future issues to understand the proper behavior of entanglement. We hope our
result would have far reaching consequences in entanglement dynamics.

\vspace{.2cm}

{\protect\( ^{1}\protect \)ichattopadhyay@yahoo.co.in}

{\protect\( ^{2}\protect \)dsappmath@caluniv.ac.in}

{\textbf{Appendix:} Consider two pure bipartite states
$|\Psi\rangle, |\Phi\rangle$ of Schmidt rank $d$ having Schmidt
vectors, $(\alpha_1, \alpha_2, \cdots \alpha_d),$ and $(\beta_1,
\beta_2, \cdots \beta_d)$ respectively, where $1>\alpha_i \geq \alpha_{i+1}>0$ and
$1>\beta_i \geq \beta_{i+1}>0,~\forall~~i=1,2, \ldots,
d-1,~~\sum_{i=1}^d \alpha_i = 1= \sum_{i=1}^d \beta_i$. Suppose it
is possible to convert $|\Psi\rangle$ to $|\Phi\rangle$ under
deterministic LOCC, i.e., $|\Psi\rangle \rightarrow |\Phi\rangle$.
Then, from Nielsen's criteria we have,
\begin{equation}
\begin{array}{lcl}
\sum_{i=1}^k \alpha_i \leq \sum_{i=1}^k \beta_i~~;
~~~\forall~~k=1,2, \ldots, d-1
\end{array}
\end{equation}
The above relation could be restated as,
\begin{equation}
\begin{array}{lcl}
\sum_{i=1}^k \alpha_i = \sum_{i=1}^k \beta_i -\epsilon_k; ~ \epsilon_k \geq 0, ~\forall~k=1,2, \ldots, d-1\\
\end{array}
\end{equation}

Now, for the case $\epsilon_1>\epsilon_2>\cdots>\epsilon_{d-1}>0$, we have,

$E(|\Psi\rangle)- E(|\Phi\rangle)$

$= \{-\sum_{i=1}^d
\alpha_i\log_2{\alpha_i}\} -\{- \sum_{i=1}^d \beta_i\log_2{\beta_i}\}$

$= \beta_1\log_2{\beta_1}- (\beta_1 -\epsilon_1)\log_2(\beta_1
-\epsilon_1)$

$~~ +
\sum_{i=2}^{d-1} \{\beta_i\log_2{\beta_i}- (\beta_i
+\epsilon_{i-1}-\epsilon_i)\log_2(\beta_i
+\epsilon_{i-1}-\epsilon_i)\}$

$~~~+ \beta_d\log_2{\beta_d}- (\beta_d +\epsilon_{d-1})\log_2(\beta_d
+\epsilon_{d-1})$

$= -\beta_1\log_2(1 -\frac{\epsilon_1}{\beta_1})- \sum_{i=2}^{d-1}
\beta_i\log_2(1
+\frac{\epsilon_{i-1}-\epsilon_i}{\beta_i})$

$~~~- \beta_d \log_2(1+\frac{\epsilon_{d-1}}{\beta_d})+ \epsilon_1\log_2\alpha_1+
\sum_{i=2}^{d-1}(\epsilon_i-\epsilon_{i-1})\log_2\alpha_i$

$ ~~~~-\epsilon_{d-1}\log_2\alpha_d$

$\geq \log_e(2) (\beta_1 \sum_{j=1}^\infty
\frac{(\frac{\epsilon_1}{\beta_1})^j}{j} + \sum_{i=2}^{d-1} \beta_i
(\frac{\epsilon_{i-1}-\epsilon_i}{\beta_i})$

$~~~+\beta_d (\frac{\epsilon_{d-1}}{\beta_d}))
+ \sum_{i=1}^{d-1}\epsilon_i(\log_2\alpha_{i}-\log_2\alpha_{i+1})$

$= \log_e(2) (\epsilon_1 + \beta_1 \sum_{j=2}^\infty
\frac{(\frac{\epsilon_1}{\beta_1})^j}{j} - \sum_{i=2}^{d-1}
(\epsilon_{i-1}-\epsilon_i) + \epsilon_{d-1})$

$~~~~~~+ \sum_{i=1}^{d-1}\epsilon_i\log_2(\frac{\alpha_i}{\alpha_{i+1}})$

$= \log_e(2) ( \beta_1 \sum_{j=2}^\infty
\frac{(\frac{\epsilon_1}{\beta_1})^j}{j})
+ \sum_{i=1}^{d-1}\epsilon_i\log_2(\frac{\alpha_i}{\alpha_{i-1}})
>0.$

The proof of all other alternatives are similar.

\end{document}